\documentclass{aa}
\usepackage{graphicx}
\usepackage{natbib}
\bibpunct{(}{)}{;}{a}{}{,}

\begin{document}

   \title{The Composition of the Interstellar Medium towards the Lockman Hole}

   \subtitle{\ion{H}{i}, UV and X-ray observations}

   \author{M. Kappes\inst{1}, J. Kerp\inst{1} \and P. Richter\inst{2}}

   \offprints{M. Kappes}

   \institute{Radioastronomisches Institut der Universit\"at Bonn, Auf dem H\"ugel 71, 53121 Bonn, Germany\\
              \email{mkappes@astro.uni-bonn.de, jkerp@astro.uni-bonn.de} 
              \and Osservatorio Astrofisico di Arcetri, Largo E. Fermi, 5, 50125 Florence, Italy\\
              \email{richter@arcetri.astro.it}}

   \date{Received 2002; accepted 2002}

   \abstract{
The Lockman Hole is well known as the region with the lowest neutral atomic hydrogen column
density on the entire sky. We present an analysis of the soft X-ray background 
radiation towards the Lockman Hole using ROSAT all-sky survey data. This data is correlated
with the Leiden/Dwingeloo survey (Galactic \ion{H}{i} 21\,cm-line emission) in order
to model the soft X-ray background by using radiative transfer calculations for four
ROSAT energy bands simultaneously. It turns out, that an important gas fraction,
ranging between 20--50\%,
of the X-ray absorbing material is not entirely traced by the \ion{H}{i} but is in the
form of ionized hydrogen. Far-ultraviolet absorption line measurements by FUSE are consistent
with this finding and support an ionized hydrogen component towards the Lockman Hole.

   }
\titlerunning{The ISM towards the Lockman Hole}
   \maketitle

%________________________________________________________________
\section{Introduction}
\label{introduction}
In this paper we focus on
the Galactic plasma emission with emphasis on the coronal gas in the Milky Way halo.
We concentrate our study on the Lockman Hole region on the northern high
Galactic latitude sky. The Lockman Hole (e.g. \citet[][]{loc86}, \citet[][]{jah90}) is the sky region 
with the absolute lowest
\ion{H}{i} column density ($N_{\ion{H}{i}}\approx 4.5 \cdot 10^{19}\,{\rm cm}^{-2}$). Therefore, it is 
considered to be the most transparent window to the distant Universe. 
To test this assumption, we investigate the photoelectric absorption of X-rays by the interstellar medium (ISM).

Towards this aim, we performed a correlation analysis of the ROSAT all-sky survey data
with the Leiden/Dwingeloo \ion{H}{i} 21\,cm-line survey. Both surveys are the
state-of-the-art data bases for such an approach. In the soft X-ray energy regime (E $<$ 1\,keV)
the mean free path length of the X-ray photons are only a few tens of parsec at maximum, assuming a
density of $n \simeq 1\,{\rm cm}^{-3}$ \citep[][]{mcc90}.
The exact knowledge of the amount of photoelectric absorbing material is essential
to evaluate the soft X-ray radiative transfer through the Galactic ISM.
Previous investigations of the Lockman Hole area \citep[][]{sno94a} showed that diffuse soft X-ray 
emission originates beyond some \ion{H}{i} gas clouds on the line of sight.
They concluded that in general the 1/4\,keV X-ray emission appears to be anti-correlated 
with the $N_{\ion{H}{i}}$. In detail, significant deviations between the localization of individual 
\ion{H}{i} clouds and the X-ray intensity minima are suggested. Moreover, \citet[][]{sno94a} 
proposed that the X-ray emitting halo gas located beyond the Draco cloud \citep[][]{sno91} and 
the one observed towards the Lockman Hole are different in temperature $T$ and emission measure 
$n_{\rm e}^2$ indicating a different origin of both plasmas. This supports the picture of a patchy 
Milky Way halo gas.

In this paper, we analyze a much larger portion -- enclosing the Lockman Hole -- of the northern 
high Galactic latitude sky. We use the ROSAT all-sky survey data of four energy bands, sampling 
the soft energy regime: 0.25\,keV  $\leq E \leq$ 0.82\,keV. The Leiden/Dwingeloo \ion{H}{i} 
survey data serves as a tracer for the photoelectric absorption, evaluated by the \ion{H}{i} column 
density along the lines of sight. We utilize the velocity information inherent to the \ion{H}{i} 
line spectra to unravel spatially the composition of the ISM towards the Lockman Hole. In particular, 
we focus on the role of the Warm Neutral Medium (WNM) and the intermediate-velocity cloud gas (IVC) towards 
this position on the sky. 

The paper is organized as follows:
Section 2 presents the X-ray and \ion{H}{i} data.
Section 3 outlines our approach to correlate both data sets, summarizes the 
best fit parameters for the temperatures, intensities and absorbing column densities.
Section 4 describes the far-ultraviolet absorption line analysis of the Lockman Hole.
In Section 5 we present our conclusions.

\section{The Data}
\subsection{\ion{H}{i} Data}
\label{hidata}
The Leiden/Dwingeloo survey of Galactic neutral hydrogen \citep[][]{har97} covers the 
entire sky north of $\delta = -30^{\circ}$ with a true-angle spacing of 0\fdg5 in 
both {\it l} and {\it b}. The spectrometer used has 1024 channels and a velocity resolution 
of 1.03\,km\,s$^{-1}$ which leads to a LSR velocity coverage of {-450}\,km\,s$^{-1}$ to +400\,km\,s$^{-1}$.
Hence, most high-velocity cloud (HVC) emission is encompassed by this velocity interval. 
The rms-noise of the measured brightness-temperature amounts to $\Delta T_{\rm B}=0.07$\,K.
Moreover, the spectra are corrected for the influence of stray radiation \citep[][]{kal80, har96}, which 
is of greatest concern in galactic \ion{H}{i}-surveys. 

We extracted the sky region bordered by $b = 12^{\circ}$\dots$75^{\circ}$ and 
$l = 99^{\circ}$\dots$166^{\circ}$ including the Lockman Hole. The lowest \ion{H}{i} column density 
region \citep[][ ($N_{\ion{H}{i}} \approx 4.5 \cdot 10^{19}\,{\rm cm}^{-2}$)]{loc86} is localized 
at $(l,b) = (150^{\circ}, 53^{\circ})$. 
According to the large stretch in Galactic latitude, the contrast between the highest column density region at 
low Galactic latitude and the Lockman Hole is about a factor of 25 (equivalent $\simeq$ 14 dB). This high column density 
contrast gives us the opportunity to disentangle the contribution of the different diffuse X-ray emission 
components, by focussing on regions where each individual X-ray emission component is observable with
high significance. The \ion{H}{i} data are binned into velocity integrated maps which are projected on the 
ROSAT grid for a pixel-by-pixel correlation. This is described in Sect. \ref{modeling}. The pixel size is set 
to 48\arcmin, about twice the ROSAT all-sky survey data analyzed here and about $\sqrt{2}$ of the Leiden/Dwingeloo 
beam size.

\subsection{X-ray Data}
\label{xdata}
The X-ray data of the Lockman area were extracted from the ROSAT all-sky survey (RASS) \citep[][]{sno95, sno97}. 
The data used have an angular resolution of 24\arcmin. We investigate four ROSAT energy bands: the R1-band 
($E \simeq 0.25$\,keV), the R2-band ($E \simeq 0.31$\,keV), the C-band ($E \simeq 0.28$\,keV), and the 
M-band ($E \simeq 0.82$\,keV) \citep[][]{sno94}. The data are corrected for scattered solar X-rays \citep[][]{sno93}, 
particle background \citep[][]{plu93}, as well as long--term X-ray enhancements \citep[][]{sno95}.
X-ray point sources were removed down to a minimum count rate of 0.02\,cts\,s$^{-1}$ \citep[][]{sno97}.
The statistical significance of the X-ray data were evaluated by the ROSAT uncertainty maps which only 
account for the number of photon events: they do not include any systematic uncertainties produced by any non-cosmic 
X-ray events.

%________________________________________________________________
\section{Method}
\subsection{Radiative transfer equation for the ISM}
\label{radiation}
\subsubsection{The effective photoelectric absorption}
\label{effphot}
In the soft X-ray energy regime, the photoelectric absorption is the dominant process of X-ray attenuation.
According to \citet[][]{mor83}, the photoelectric absorption cross section is a strong function of 
energy ($\sigma \propto E^{-\frac{8}{3}}$) and amounts to some $\sigma \simeq 10^{-20}\,{\rm cm}^2$ 
(according to ROSAT C-band) for an ISM of 
solar metalicity and $n \simeq 1\,\mathrm{cm^{-3}}$.
Both findings highlight two important facts: a) the lower the X-ray photon energy the higher the probability of
photoelectric absorption and b) all lines of sight through the Galactic ISM are optically thick for soft 
X-ray emission -- except for the Lockman Hole. As pointed out in Sect. \ref{xdata}, we analyze the X-ray 
intensity distribution of broad energy bands. According to this approach, it is necessary to evaluate the 
X-ray band--averaged photoelectric absorption cross section, the so-called {\em effective photoelectric 
absorption cross section}:
\begin{equation}
\sigma(N_{\ion{H}{i}}) = -\frac{1}{N_{\ion{H}{i}}} \ln \left\{\frac{\int S(E)\cdot T(E)\cdot {\rm e }^{-\sigma'(E)\cdot N_{\ion{H}{i}}}{\rm \cdot dE}} {\int S(E)\cdot {\rm e }^{-\sigma'(E)\cdot N_{\ion{H}{i}}}{\rm \cdot dE}}\right\}
\label{eqn_effcs}
\end{equation}
Here $S(E)$ denotes the spectrum of the X-ray source, $T(E)$ the telescope and detector response function, 
$\sigma'(E)$ the \citet[][]{mor83} photoelectric cross section, $N_{\ion{H}{i}}$ the observed \ion{H}{i} column density
 \citep[for details see][]{ker99}.
Using this effective photoelectric absorption cross section $\sigma(N_{\ion{H}{i}})$ yields an inherent ambiguity:
without an {\em educated guess} of $S(E)$ it is not possible to determine $\sigma(N_{\ion{H}{i}})$.
In consequence, we have to compare the observational data with a variety of different spectral models to overcome 
this ambiguity.

\subsubsection{Galactic $N_{\ion{H}{i}}$}
\label{galhi}
The distribution of the general ISM of the Milky Way is traced by the Galactic \ion{H}{i} 21\,cm-line emission. 
It is thought that \ion{H}{i} -- as the most abundant element in space -- traces the distribution of the metals 
quantitatively.

Early \ion{H}{i} surveys already revealed that \ion{H}{i} clouds exist next to the rotating Galactic disk, which 
cannot be modeled within the framework of co-rotating ISM.
Based on these early findings, today we distinguish between low-, intermediate- and 
high-velocity clouds \citep[][]{dic90}. 
In the following we will show, that low- and intermediate-velocity clouds can be identified as soft X-ray shadows against 
the distant diffuse X-ray background emission (e.g. \citet[][]{sno00}). 
To differentiate between these different constituents of the ISM we 
bin the data into velocity integrated maps covering 
$|v_\mathrm{LSR}| < 25\,{\rm km\,s^{-1}}$, 
$|v_\mathrm{LSR}| < 75\,{\rm km\,s^{-1}}$ and 
$|v_\mathrm{LSR}| < 100\,{\rm km\,s^{-1}}$. We show that the $|v_\mathrm{LSR}| < 75\,{\rm km\,s^{-1}}$ already 
covers \ion{H}{i} gas belonging
to the M81/M82 group of galaxies. This intra-group \ion{H}{i} gas allows a critical test of our X-ray vs. \ion{H}{i} 
correlation analysis because it is accidentally included in the radiative transfer equation
as an absorber within the $|v_\mathrm{LSR}| < 75\,{\rm km\,s^{-1}}$ map of the Milky Way halo. 

\subsection{The individual X-ray source terms}
\subsubsection{The Local Hot Bubble}
Early sounding rocket experiments as well as the Wisconsin X-ray sky survey established the existence of hot X-ray 
emitting plasma within the local environment of the Sun. This coronal gas fills the irregularly shaped local void 
of matter \citep[][]{mcc90} -- frequently called the Local Hot Bubble (LHB). Therefore, the X-ray intensity of the 
Local Hot Bubble varies appreciable across the entire sky: 
$I_{\rm LHB} = (2.5 \dots 8.2)\cdot 10^{-4}\,{\rm cts\,s^{-1}\,arcmin^{-2}}$ \citep[][]{sno98}. 
$I_{\rm LHB}$ is the first term in our radiative transport equation (see Eqn.\,\ref{eqn_radiation}). Considering the 
results of \citet[][]{sfe99} we 
know that only a few local clouds are intermixed with the hot X-ray emitting plasma of the Local Hot Bubble 
\citep[e.g.][]{ker93}. Because of this finding, we assume that $I_{\rm LHB}$ is not absorbed by any intervening gas 
along the line of sight.

\subsubsection{The Milky Way halo}
The existence of hot coronal gas within the Milky Way halo was one of the major discoveries of the ROSAT mission.
Moreover, it was found that the X-ray emission from the LHB has a superposed diffuse component 
\citep[][]{sno91,her95,ker96,pie98} in excess to the contribution of the extragalactic background radiation.
It is beyond any doubt that diffuse X-ray emission originates within the Milky Way halo, but it is a matter of debate 
as to how many thermal plasma components are localized within the Galactic halo \citep[][]{kun00}. Their corresponding 
individual temperatures and emission measures are also a matter of debate. Here we follow the approach of a hot single 
halo component which will turn out to be sufficient. We vary the 
temperature of the Galactic halo plasma and search for a best fit temperature simultaneously consistent with all four 
ROSAT energy band data.
The Milky Way halo is represented as $I_{\rm HALO}$ in the radiative transfer equation (see Eqn.\,\ref{eqn_radiation}). 
This diffuse X-ray emission 
component is absorbed by the Galactic ISM along the line of sight. Following the latest results on the scale height of
the highly ionized species \citep[][]{sav97,sav03} and prior \ion{H}{i} vs. X-ray correlation analysis \citep[][]{pie98}, the 
scale height of the Milky Way halo plasma is about 4\,kpc \citep[][]{kal98}. We can safely assume that all ISM 
components with a lower scale height, the Warm Neutral Medium (WNM), the Warm Ionized Medium (WIM) and the Cold Neutral 
Medium (CNM) are 
in front of this diffuse X-ray emission. This amount of absorbing material has to be evaluated in order to calculate the 
strength of the photoelectric absorption. We account for the different characteristic scale heights of the neutral species
by using the velocity integrated maps defined in Sect. \ref{galhi}.

\subsubsection{The extragalactic component}
The extragalactic X-ray background is composed of superposed emission of point sources, in particular of 
AGNs \citep[e.g.][]{toz01}. Its spectrum can be approximated by a power law ($E^{- \Gamma}$). We assume an average 
spectral index of $\Gamma = 1.5$ according to \citet{has01}. Its ROSAT C-band intensity was determined by \citet{bar96}; 
they derived $I_{\rm EXTRA} = 228 \cdot 10^{-6}\,{\rm cts\,s^{-1}\,arcmin^{-2}}$ with an uncertainty of 
$90 \cdot 10^{-6}\,{\rm cts\,s^{-1}\,arcmin^{-2}}$. Throughout the fit procedure (see Sect. \ref{modeling}) we will 
fix $I_{\rm EXTRA}$ to both values normalized to the ROSAT C-band and their corresponding values in the other ROSAT 
energy bands respectively. We assume that {\em all} neutral interstellar matter along the line of sight measured in the 
Leiden/Dwingeloo \ion{H}{i} survey ($N_{\ion{H}{i},\rm e}$) is in front of the extragalactic sources. Also this amount of 
photoelectric absorbing material is constant in the evaluation of the soft X-ray radiative transfer calculations.

This leads to the radiative transfer equation for soft X-rays introduced by \citet{ker99}: 
\begin{equation}
I = I_{\rm LHB} + I_{\rm HALO} \cdot {\rm e}^{- \sigma_\mathrm{h} \cdot N_{\ion{H}{i},{\rm h}}} + I_{\rm EXTRA} \cdot {\rm e}^{- \sigma_\mathrm{e} \cdot N_{\ion{H}{i},{\rm e}}}.
\label{eqn_radiation}
\end{equation}

Here, $\sigma_{\rm h}$ denotes the effective photoelectric absorption cross section of the Galactic halo, while 
$\sigma_\mathrm{e}$ represents that for the extragalactic background. While we can safely assume that all \ion{H}{i} 
observed towards a distant AGN is enclosed within the velocity range 
$-450\,{\rm km\,s^{-1}}<\,v_\mathrm{LSR}\,<450\,{\rm km\,s^{-1}}$, we 
have to question this assumption with respect to the Milky Way. Here we have to identify the best fit velocity range by 
comparing the \ion{H}{i} and the diffuse X-ray data quantitatively. To differentiate between both absorbing column 
densities we use $N_{\ion{H}{i},{\rm e}}$ and $N_{\ion{H}{i},{\rm h}}$ from here on. Equation \ref{eqn_radiation} 
has {\em four} free parameters ($I_{\rm LHB}, I_{\rm HALO}, T_{\rm LHB}, T_{\rm HALO}$) and {\em two} fixed 
parameters ($I_{\rm EXTRA}, \Gamma$). The temperatures of the LHB ($T_{\rm LHB}$) and of the 
Galactic halo ($T_{\rm HALO}$) are connected to Eq. \ref{eqn_radiation} by the source term $S(E)$ in Eq. \ref{eqn_effcs}.
$N_{\ion{H}{i},{\rm e}}$ and $N_{\ion{H}{i},{\rm h}}$ are fixed parameters as 
well because by optimizing the free parameters both $N_{\ion{H}{i}}$ values are fixed. To determine the four free 
parameters, we use the information of the ROSAT R1-, R2-, C- and M-Band in {\em simultaneous} fit procedures of 
X-ray energy band ratios and intensities vs. $N_{\ion{H}{i},{\rm h}}$.

\subsection{Modeling}
\label{modeling}
In order to evaluate the radiative transport equation (Eq. \ref{eqn_radiation}) for all ROSAT energy bands
and band ratios simultaneously, we developed a two-step fit procedure. The first step ({\em one dimensional approach}) 
gives us the opportunity to estimate the start-values for the fit parameters. The second step ({\em two dimensional 
approach}) is used to derive the spatial distribution of the X-ray intensities correctly. Moreover, we can optimize the
fit parameters within the framework of the two dimensional fit procedure.

\subsubsection{The one dimensional approach}
\label{onedim}
\begin{figure}[htb]
\centering
\includegraphics[width=7cm, angle=0]{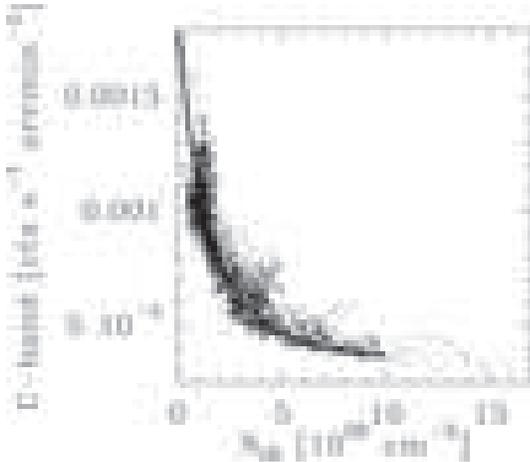}
\caption{Scatter plot for the ROSAT C-band. At high column densities ($>5 \cdot 10^{20}$\,cm$^{-2}$) the X-ray
intensity shows an asymptotic behavior. This represents only the foreground emission of the LHB.
The solid line shows the best fit model which satisfies {\em all} ROSAT energy bands and band ratios simultanously
(see Fig. \ref{fig_s_m}, \ref{fig_s_c2m}, \ref{fig_s_r1r2} as well). 
Other models (i.e. with varying halo- and/or LHB-temperatures) are not distinguishable from each other.}
\label{fig_s_c}
\end{figure}

The radiative transport equation (see Eq. \ref{eqn_radiation}) gives us the possibility to derive fit parameters, 
namely the intensities of the two free emission components. In practice we calculate so--called {\em scatter diagrams} 
(see Fig.\,\ref{fig_s_c} for the C-band scatter diagram) for the ROSAT C- and M-band and the energy band ratios R1/R2 
and C/M as functions of $N_\ion{H}{i}$. The energy band ratios are a sensitive measure for the temperature of 
both X-ray emitting plasmas.

\begin{figure}[htb]
\centering
\includegraphics[width=7cm, angle=0]{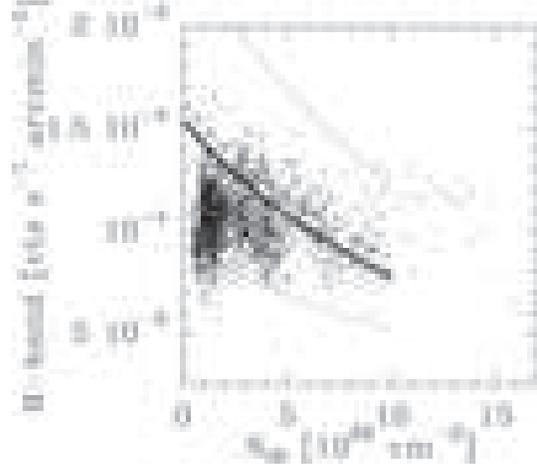}
\caption{Scatter plot for the ROSAT M-band. The dotted lines correspond to models where the halo plasma
temperature is $10^{6.3}$\,K (upper curve) and $10^{6.1}$\,K (lower curve). The dashed lines correspond
to models where the LHB temperature is $10^{5.9}$\,K (upper curve) and $10^{6.1}$\,K (lower curve). As 
in Fig.\,\ref{fig_s_c} the solid line represents the best fit model.}
\label{fig_s_m}
\end{figure}

The first free parameter value we derive is the C-band intensity of the Local Hot Bubble ($I_{\rm LHB}$).
The data points in the C-band scatter diagram (Fig.\,\ref{fig_s_c}) show an asymptotic behavior at high 
column densities 
($> 5 \cdot 10^{20}\,{\rm cm^{-2}}$). Here we observe only the X-ray foreground emission because the entire background 
emission of $I_\mathrm{HALO}$ and $I_\mathrm{EXTRA}$ is absorbed by the high column density in front of both components;
therefore, the first estimate is $I_{\rm LHB} = 450 \cdot 10^{-6}\,{\rm cts\,s^{-1}\,arcmin^{-2}}$ which is consistent 
with the values given by \citet{sno98}.

The $I_{\rm HALO}$ is determined at the low column density part of the C-band and full M-band scatter diagram.
In the ROSAT C-band we observe the superposed emission of $I_{\rm LHB}$ and $I_{\rm HALO}$ while in the ROSAT 
M-band (see Fig.\,\ref{fig_s_m}) we have to include the diffuse X-ray emission attributed to the extragalactic background.

\begin{figure}[htb]
\centering
\includegraphics[width=6cm, angle=0]{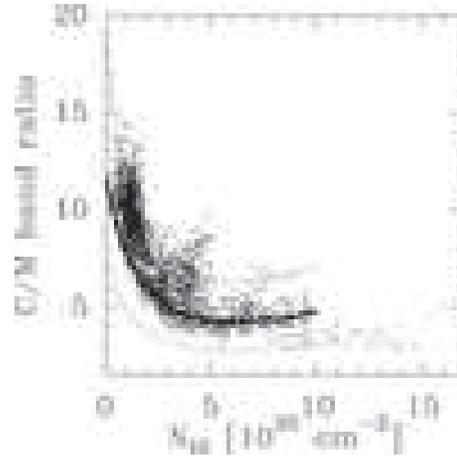}
\caption{ROSAT C/M-band ratio vs. $N_{\ion{H}{i}}$. The dotted lines correspond to models where the halo plasma
temperature is $10^{6.3}$\,K (upper curve) and $10^{6.1}$\,K (lower curve). As 
in Fig.\,\ref{fig_s_c} the solid line represents the best fit model. Note, that the dashed lines, representing the
variation of $T_{\rm LHB}$ in Fig. \ref{fig_s_r1r2}, are not suitable to evaluate $T_{\rm LHB}$ in the C/M-band ratio.}
\label{fig_s_c2m}
\end{figure}

\begin{figure}[htb]
\centering
\includegraphics[width=6cm, angle=0]{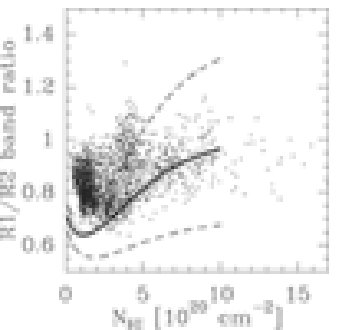}
\caption{ROSAT R1/R2-band ratio vs. $N_{\ion{H}{i}}$. The dashed lines correspond
to models where the LHB temperature is $10^{5.9}$\,K (upper curve) and $10^{6.1}$\,K (lower curve). As 
in Fig.\,\ref{fig_s_c} the solid line represents the best fit model. Note, that the dotted lines, representing the
variation of $T_{\rm HALO}$ in Fig. \ref{fig_s_c2m}, are not sensitive to evaluate $T_{\rm HALO}$ in the C/M-band ratio.}
\label{fig_s_r1r2}
\end{figure}

Considering only the C-band scatter diagram (Fig.\,\ref{fig_s_c}), an ambiguity concerning the plasma 
temperatures remains. Variations of the plasma temperature $T_\mathrm{HALO}$ yield insignificant variations in 
the fitted curves. To constrain $T_\mathrm{LHB}$ and $T_\mathrm{HALO}$ we plot the R1/R2 and C/B band ratio versus 
$N_\mathrm{HI,h}$ (see Fig.\,\ref{fig_s_c2m} \& \ref{fig_s_r1r2}). Regarding these, we see large variations in the 
C/M Band ratio vs. $N_\mathrm{HI,h}$ plot, when we vary the halo plasma temperature in the same way as for the 
C-band scatter diagram (see Fig.\,\ref{fig_s_c}). Variations of $T_\mathrm{LHB}$ cause significant 
variations in the R1/R2 band ratio vs. $N_\mathrm{HI,h}$ (Fig.\,\ref{fig_s_r1r2}). An immediate impact on 
the energy band ratios vs. $N_\mathrm{HI,h}$ is seen if we vary $I_{\rm HALO}$ and $I_{\rm LHB}$ individually. 
Because a variation of only one free parameter has an immediate effect, all four scatter diagrams, we have a 
sensitive tool to find a consistent solution for all four parameters simultaneously. To illustrate this proposal 
we plot a set of temperature models for $T_\mathrm{HALO}$  in each of the four scatter diagrams while leaving 
the other parameters $I_\mathrm{LHB}$, $I_\mathrm{HALO}$ and $T_\mathrm{LHB}$ fixed to the best fit parameters 
given in Sec. \ref{optimize}. Since the C-band and R1/R2 band ratio scatter diagrams are not sensitive to temperature 
variations of the parameter $T_\mathrm{HALO}$ the M-band and C/M band ratio scatter diagrams show up with significant 
deviations from the best fit situation marked by the solid line.

In Fig.\,\ref{fig_s_r1r2} we demonstrate the variation of the $T_\mathrm{LHB}$ on the R1/R2 band ratio scatter diagram. 
While the C- and M-band scatter diagrams as well as the C/M ratio scatter diagram are not sensitive measures for 
$T_\mathrm{LHB}$ we find that the R1/R2 band ratio scatter diagram allows one to determine $T_\mathrm{LHB}$ very 
accurately.

The one-dimensional approach performed here indicates that the field--averaged values of the four free parameters can 
be accurately determined simultaneously. The R1/R2 band ratio scatter diagram is a measure for $T_\mathrm{LHB}$. 
The C/M ratio scatter diagram is a measure for $T_\mathrm{HALO}$. The C-band scatter diagram determines $I_\mathrm{LHB}$ 
and $I_\mathrm{HALO}$, while the M-band scatter diagram is a measure for $I_\mathrm{HALO}$ and $T_\mathrm{HALO}$.

These four parameters form the smallest set of parameters which are necessary to model the X-ray intensity distribution 
in the R1-, R2-, C- and M-band. The one dimensional approach leads to the following initial fit parameters:

\begin{eqnarray*}
I_{\rm LHB} & = & (450 \pm 90) \cdot 10^{-6}\,{\rm cts\, s^{-1}\, arcmin^{-2}} \\
I_{\rm HALO} & = & (1200 \pm 120) \cdot 10^{-6}\,{\rm cts\, s^{-1}\, arcmin^{-2}} \\
I_{\rm EXTRA} & = & (230 \pm 90) \cdot 10^{-6}\,{\rm cts\, s^{-1}\, arcmin^{-2}} \\
\\
T_{\rm LHB} & = & 10^{6.0 \pm 0.1}\,{\rm K} \\
T_{\rm HALO} & = & 10^{6.2 \pm 0.1}\,{\rm K}. 
\end{eqnarray*}

\subsection{The two dimensional approach}
\label{twodim}
Extending our analysis of the diffuse X-ray intensity distribution from one to two dimensions allows us to disclose 
systematic sources of uncertainties hidden by the scatter of the X-ray counts. Namely variations of $I_\mathrm{LHB}$ 
and $I_\mathrm{HALO}$ as functions of $l$ and $b$ on small as well as on large angular scales.
 
In our two dimensional approach we evaluate Eq. \ref{eqn_radiation} in a pixel-by-pixel manner. For simplicity we assume 
that the four parameters are constant across the whole sky region we study.

\begin{figure}[htb]
\centering
\includegraphics[width=6cm, angle=0]{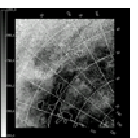}
\includegraphics[width=6cm, angle=0]{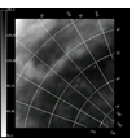}
\caption{{\bf upper panel:} The color image shows the observed ROSAT C-band while the contours correspond
to the deviation map. The solid contours encircle regions where the modeled intensities are too faint, the
dotted contours enclose regions where the model is too bright. Intra-group gas of the M81/M82 group of galaxies
within the spectra of the Leiden/Dwingeloo \ion{H}{i} survey produces the contour line 
around $(l,b)\approx (140^\circ,40^\circ)$. HVC complex C is located at the bottom, the Lockman hole is
surrounded by the dotted contour line on the right hand side.
{\bf lower panel:} The color image shows the modeled ROSAT M-band. The superimposed contours of the corresponding
deviation map are almost nonexistent, indicating a quantitatively good model within the uncertainties.
Intensities are in $10^{-6}\,$cts\,s$^{-1}$\,arcmin$^{-2}$.}
\label{deviationmap}
\end{figure}

Equipped with the initial parameters, we calculate model maps for the C- and M-band and compare them to the observed 
X-ray distribution in these bands. This is done by subtracting the model maps from the observed ones and dividing the
difference by the ROSAT uncertainty maps. The result is a {\em deviation map} which shows the discrepancy of observation 
and model in units of the statistical standard deviation. These deviation maps (Fig.\,\ref{deviationmap}) reveal that most
of the observed ROSAT C- and M-band data are fully consistent with the corresponding modeled maps. While the ROSAT M-band 
data matches with the modeled M-band (within the uncertainties), the C-band shows some large scale deviations
between the observed and the modeled X-ray distribution. Especially, the intra-group gas of the M81/M82 group of
galaxies leads to an expected deviation in the ROSAT C-band (see Fig.\,\ref{deviationmap}, upper panel). This is caused
by the admixed \ion{H}{i} line emission of the intra-group gas and the Galactic hydrogen. %(see Fig.\,\ref{hispectra}).
The supernumerary \ion{H}{i} column density is seen as an X-ray shadow in the modeled ROSAT C-band located at the
correct position on the sky. In this sense, we can test the two dimensional approach spatially.

Towards the low- and high-latitude regions of the maps the observed and the modeled maps agree 
quantitatively (Fig.\,\ref{deviationmap}). 
Here, the dynamic range in X-ray absorbing \ion{H}{i} column density reaches extreme values. The quantitative 
agreement of these areas indicates that the general approach is successful. However, the differences between 
both maps are pronounced towards the HVC-complex C \citep[][]{ker99} and the Lockman Hole. Here, we focus on 
the Lockman Hole.

\subsubsection{Testing the two dimensional correlation}
To optimize the initial fit parameters (see Sec. \ref{modeling}) we compute histograms of the deviation maps and 
obtain the mean ($\mu$) and variance ($\sigma$).
In case of a perfect quantitative correlation between the observation and the model we expect a mean of $\mu=0$ and 
a variance of $\sigma=1$. This can be achieved by small variations of the free fit parameters and slight changes 
in the \ion{H}{i} column density of the halo component (i.e. the velocity interval of the \ion{H}{i} data).
It turns out that the quality of the histograms increases when the intensity of the extragalactic background emission
is tuned to $(170 \pm 90) \cdot 10^{-6}\,{\rm cts\, s^{-1}\, arcmin^{-2}}$ which is fully consistent 
with \citet[]{bar96} within the uncertainties. After seven iteration steps no further improvement can be achieved. The 
statistical parameters in the C- and M-band after the optimization are 
$\mu_{\rm C}=0.13, \sigma_{\rm C}=1.56$ and $\mu_{\rm M}=-0.58, \sigma_{\rm M}=1.12$ respectively. The final C-band 
model map is shown in Fig.\,\ref{c_mod_70} for comparison with the observed C-band intensity 
distribution (Fig.\,\ref{deviationmap}).

\subsubsection{The role of $N_{\ion{H}{i}}$}
\label{optimize}
\begin{figure}[htb]
\centering
\includegraphics[width=6cm, angle=0]{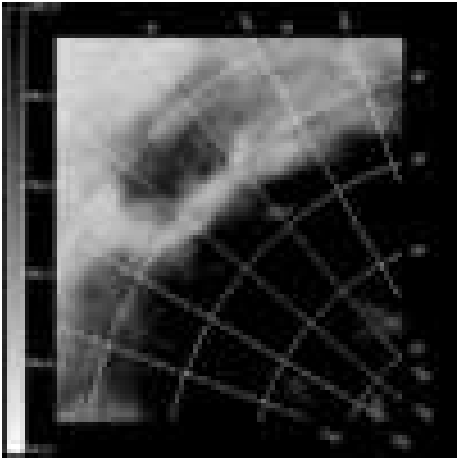}
\caption{Modeled ROSAT C-band map. The amount of \ion{H}{i} corresponds to the velocity interval 
of $|v_{\rm LSR}| \leq 25$\,km\,s$^{-1}$ which does not include the IVC regime. Therefore,
the X-ray shadow produced by IVC\,135+52-45 (compare to Fig.\,\ref{deviationmap}, upper panel) is absent.
Intensities are in $10^{-6}\,$cts\,s$^{-1}$\,arcmin$^{-2}$.}
\label{c_mod_25}
\end{figure}

\begin{figure}[htb]
\centering
\includegraphics[width=6cm, angle=0]{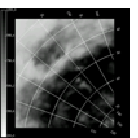}
\caption{Modeled ROSAT C-band map for the \ion{H}{i} velocity interval of $|v_{\rm LSR}| \leq 50$\,km\,s$^{-1}$.
The mean intensity is about 11\% smaller than the one of Fig.\,\ref{c_mod_25}.
Intensities are in $10^{-6}\,$cts\,s$^{-1}$\,arcmin$^{-2}$.}
\label{c_mod_50}
\end{figure}

\begin{figure}[htb]
\centering
\includegraphics[width=6cm, angle=0]{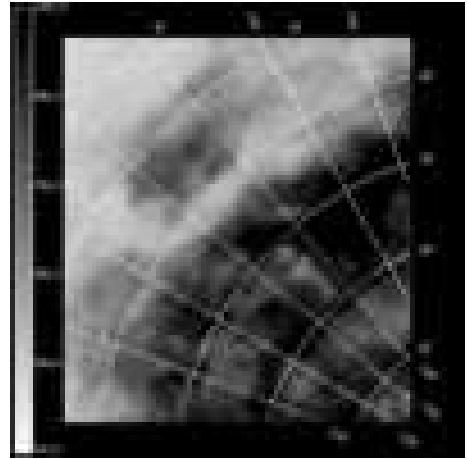}
\caption{Modeled ROSAT C-band map for the \ion{H}{i} velocity interval of $|v_{\rm LSR}| \leq 70$\,km\,s$^{-1}$.
The mean intensity is about 18\% smaller than the one of Fig.\,\ref{c_mod_25}.
Intensities are in $10^{-6}\,$cts\,s$^{-1}$\,arcmin$^{-2}$.}
\label{c_mod_70}
\end{figure}

In order to test which \ion{H}{i} components are responsible for the major absorption we use different
$N_\mathrm{HI}$ maps (see Sect. \ref{hidata}).
Obviously the smaller the velocity integration interval the lower the $N_\mathrm{HI}$ column density. However, the 
$|v_\mathrm{LSR}| \leq 25\,{\rm km\,s^{-1}}$ map does not include the IVC shadows -- most obvious for 
IVC\,135+52-45 \citep[][]{wei99}.
The X-ray {\em excess emission} close to HVC complex C is still present, as well as the {\em deficit} of attenuating 
\ion{H}{i} towards the Lockman Hole. Figures \ref{c_mod_25}, \ref{c_mod_50} and \ref{c_mod_70} show three model maps of 
the ROSAT C-band intensity
distribution. The maps differ in the amount of \ion{H}{i} in the halo component. We calculate the maps for three 
different LSR velocity intervals: $-25\,{\rm km\,s}^{-1} \leq v_{\rm LSR} \leq +25\,{\rm km\,s}^{-1}$,
$-50\,{\rm km\,s}^{-1} \leq v_{\rm LSR} \leq +50\,{\rm km\,s}^{-1}$ and
$-70\,{\rm km\,s}^{-1} \leq v_{\rm LSR} \leq +70\,{\rm km\,s}^{-1}$. The model map to the smallest velocity interval has 
a mean intensity of $I_{\rm C} = (960 \pm 360) \cdot 10^{-6}\,{\rm cts\,s^{-1}\,arcmin^{-2}}$. The model map to the 
medium velocity interval differs by 11\,\% and the model map to the largest interval by 18\,\% 
(see Fig.\,\ref{c_mod_25} -- \ref{c_mod_70}). From these values we conclude, that most of the CNM and WNM is
already included within the smallest velocity range, which is accordingly responsible for the dominant fraction of
the absorption.
Moreover, the CNM is not present at the positions of low column density \citep[][]{loc86} and therefore the WNM 
dominates the absorption of the X-ray radiation.
This statistical approach leads to the optimized fit parameters:

\begin{eqnarray*}
I_{\rm LHB} & = & (350 \pm 90) \cdot 10^{-6}\,{\rm cts\, s^{-1}\, arcmin^{-2}} \\
I_{\rm HALO} & = & (1380 \pm 120) \cdot 10^{-6}\,{\rm cts\, s^{-1}\, arcmin^{-2}} \\
I_{\rm EXTRA} & = & (170 \pm 90) \cdot 10^{-6}\,{\rm cts\, s^{-1}\, arcmin^{-2}} \\
\\
T_{\rm LHB} & = & 10^{6.0 \pm 0.1}\,{\rm K} \\
T_{\rm HALO} & = & 10^{6.2 \pm 0.1}\,{\rm K}. 
\end{eqnarray*}

In comparison to \citet[][]{kun00} we only need one single X-ray emitting component for fitting the Galactic
halo emission. The temperature we derive is in-between the temperature of the soft and hard component 
derived by \citet[][]{kun00}. The temperature for the LHB component is slightly lower in our case.

It turns out that the entire region around the Lockman Hole $(l,b) = (150^{\circ}, 53^{\circ})$ is correctly modeled
down to the data uncertainties. 
We can account for the X-ray intensity distribution on a large sky region ($60^{\circ} \times 60^{\circ}$) whereas the 
Lockman Hole itself is too bright in the model. The question why the model first fails at this prominent position will 
be discussed in the forthcoming section.

One of the main results is that $T_\mathrm{LHB} < T_\mathrm{HALO}$. This finding is obvious in the R1/R2 band ratio 
vs. $N_\mathrm{HI,h}$. Because of the strong dependence of the photoelectric absorption on energy 
($\sigma \propto E^{-3}$), the soft X-ray photons are much stronger attenuated than the hard ones, leading to an 
apparent hardening of the X-ray 
spectrum proportional to $N_\mathrm{HI}$. This implies that the higher $N_\mathrm{HI,h}$ the lower the R1/R2 band ratio of
the $I_{\rm HALO}$ source component. Figure \ref{fig_s_r1r2} shows the contrary behavior. The higher 
$N_\mathrm{HI,h}$ the larger the R1/R2 band ratio. This finding can only be explained by a cooler LHB plasma with respect 
to the Galactic halo gas. Using the best fit plasma temperatures we derive an important contribution of the Galactic halo 
plasma to the observed M-band intensities. The $I_{\rm HALO}$ is about 4 times brighter than $I_{\rm LHB}$ within the 
ROSAT C-band, while $I_{\rm HALO} \simeq \frac{4}{3}I_\mathrm{EXTRA}$ in the ROSAT M-band. $I_{\rm HALO}$ is the dominant 
source in the ROSAT C- and M-band. We assumed the intensity of the Galactic halo plasma to be constant across
the field of interest which is consistent within the uncertainties. 
The derived fluctuations in halo intensity suggested by \citet[]{sno94a} may be due to the \ion{H}{i} velocity 
intervals, which differ from our choice. These fluctuations do not appear in our model, 
since we choose the \ion{H}{i} velocity intervals as described above.

%________________________________________________________________
\section{Specifying the ISM conditions in the Lockman Hole}
\label{LMH}
\subsection{Molecular or ionized hydrogen?}
As we pointed out in the prior sections, most of the sky region investigated can be explained by our model. The deviating
portions of the sky decompose into two types: regions modeled too faint and those modeled too bright. Here we 
concentrate on the Lockman Hole, which belongs to the latter case (see Fig.\,\ref{deviationmap}, upper panel).
This is due to a lack of absorbing material. The {\em additional} hydrogen column density required to 
remove the deviation can be estimated to $N_\ion{H}{i}^{\rm add} \approx (5 \pm 3) \cdot 10^{19}\,{\rm cm^{-2}}$. 

Since most of the sky region is properly fit by our model, we argue that the ISM towards the Lockman Hole has a different
composition to the surroundings. Either the ISM at this particular position contains a molecular gas phase
or an ionized gas phase, both of which are detectable as X-ray shadows but not quantitatively traced by the \ion{H}{i} 
data. The \ion{H}{i} column density inside the 
bordering contour lines in Fig.\,\ref{deviationmap} (upper panel) is less than $10^{20}\,{\rm cm^{-2}}$. If we
consider a molecular gas phase, a minimum \ion{H}{i} column density is needed to shield this molecular phase.
\citet{her93a}, for example, claimed that an \ion{H}{i} column density of at least $3.7 \cdot 10^{20}\,{\rm cm^{-2}}$ 
is needed to allow the formation of molecular hydrogen. From this, we conclude that a molecular gas phase is unlikely
to exist. The remaining
option is an ionized state of the hydrogen towards the Lockman
Hole. Previous studies of the interstellar $H_{\rm \alpha}$ 
emission towards the Lockman Hole \citep[][]{hau02} yield $N_\ion{H}{ii} \approx 2 \cdot 10^{19}\,{\rm cm^{-2}}$
which is of the same order of magnitude as $N_\ion{H}{i}^{\rm add}$. 
We perform a cross-check of the ionization structure towards the Lockman Hole via far-ultraviolet absorption line
measurements by FUSE.

\subsection{Far-Ultraviolet absorption line data}
\label{uv}
In order to investigate the physical conditions of the interstellar medium in the direction
of the Lockman Hole we make use of Far-Ultraviolet (FUV) absorption line data. The FUV range 
bluewards of the Lyman $\alpha$ line and redwards of the Lyman break provides a wealth of
spectral diagnostics to study interstellar gas in its various phases (e.g., \citet{sav03}; \citet{ric01a}).
A large number of FUV absorption line spectra for almost every direction in the sky is publicly available from the
MAST data archive of the {\it Far Ultraviolet Spectroscopic Explorer} ({\tt http://archive.stsci.edu/fuse/}). 
For the purpose of studying the Lockman Hole field extragalactic background sources such as QSOs and AGNs are 
favorable in order to ensure a correct continuum placement for the various absorption features, and to include
Milky Way halo gas that may contain a significant fraction of the X-ray absorbing neutral and ionized gas
in this general direction.

\subsection{NGC 3310}
From the extragalactic sight lines in the general direction $l\sim 150^{\circ}$, $b\sim 50^{\circ}$ for which
good S/N FUSE data is publicly available, we have chosen the one closest to the Lockman Hole:
NGC\,3310 ($l=156.6^{\circ}$, $b=54.1^{\circ}$). The raw data was extracted using the standard FUSE data reduction 
pipeline (CALFUSE v.2.1.6), and was then further reduced in a way similar to the data processing procedure described 
in \citet{wak03}. The velocity calibration was made by comparing absorption lines from various atomic and
molecular lines with existing \ion{H}{i} 21\,cm-line data (see below). The FUSE data have an average signal-to-noise ratio
(S/N) of $\sim 14$ per resolution element, and a spectral resolution of $\sim 20$ km\,s$^{-1}$ (FWHM).

\begin{figure}[htb]
\centering
\includegraphics[width=8cm, angle=0]{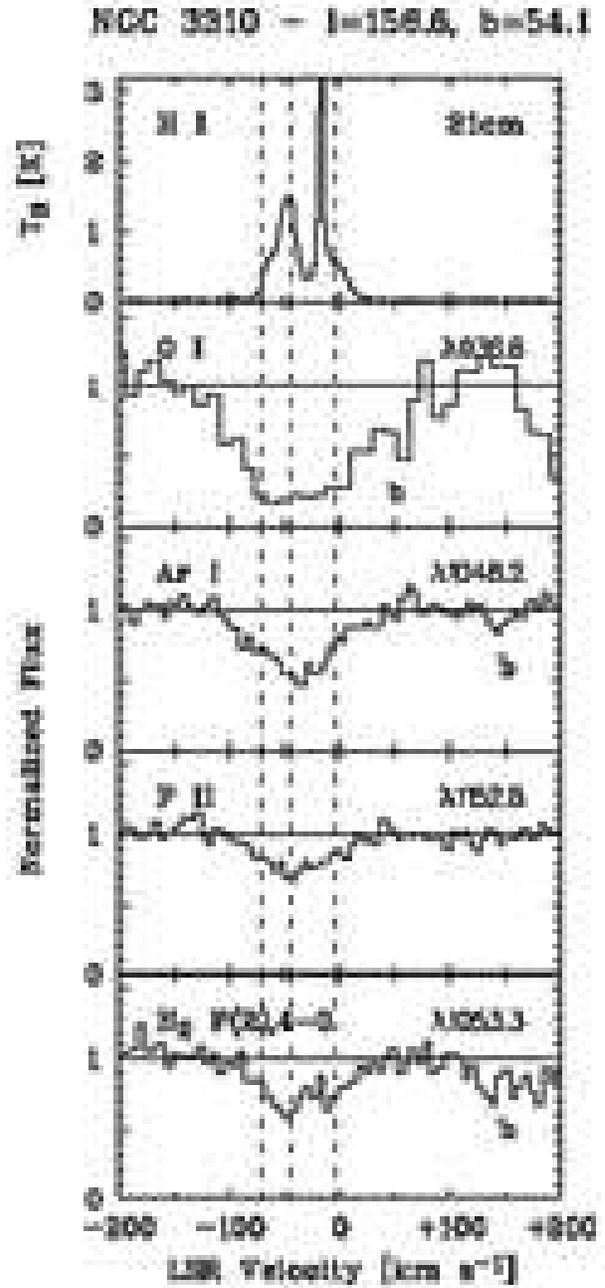}
\caption{From top to bottom, the panels show the \ion{H}{i} 21\,cm-line spectrum towards NGC\,3310 as measured by
the Green Bank Telescope, and the absorption profile of \ion{O}{i}, \ion{Ar}{i}, \ion{P}{ii}, and H$_2$ P(2),4-0 as
measured by FUSE. The dotted lines indicate the LSR velocities $-5, -45,$ and $-70$\,km\,s$^{-1}$ 
(see Sect. \ref{nwig} for details).}
\label{fig_fuse01}
\end{figure}

\begin{figure}[htb]
\centering
\includegraphics[width=8cm, angle=0]{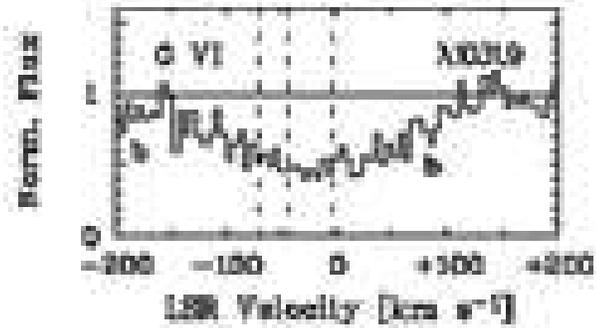}
\caption{The \ion{O}{vi} $\lambda 1031.9$ absorption profile towards NGC\,3310 traces the highly
ionized gas (see Sect. \ref{hig} for details).}
\label{fig_fuse02}
\end{figure}

\subsubsection{Neutral and weakly ionized gas}
\label{nwig}
In Fig.\,\ref{fig_fuse01} (uppermost panel) we have plotted the Green Bank \ion{H}{i} 21\,cm-line spectrum towards 
NGC\,3310 \citep[][]{mur96} on a LSR velocity scale. The 21\,cm-line spectrum shows at least four distinct 
velocity components: (1) a broad local Galactic component centered at $-11$ km\,s$^{-1}$ with 
$\log$ N($\ion{H}{i})=19.63\pm0.02$, (2) a very narrow local Galactic component at $-17$ km\,s$^{-1}$ having 
$\log$ N($\ion{H}{i})=19.54\pm0.01$, (3) a relatively strong broad component located at $-47$ km\,s$^{-1}$ with 
$\log$ N($\ion{H}{i})=19.71\pm0.01$, related to the Intermediate-Velocity Arch (IV Arch; \citet{wak01}) in the 
lower Milky Way halo, and (4) a weaker, second IV Arch component at $-65$ km\,s$^{-1}$  with
$\log$ N($\ion{H}{i})=19.11\pm0.01$. In the lower four panels of Fig.\,\ref{fig_fuse01} we have plotted the FUSE absorption
profiles of \ion{O}{i} $\lambda 936.6$, \ion{Ar}{i} $\lambda 1048.2$, \ion{P}{ii} $\lambda 1152.8$, and 
H$_2$ P(2),4-0 $\lambda 1053.3$ in direction of NGC\,3310. 

Only three velocity components (fitted by Gaussian absorption profiles) are visible in absorption,
centered at approximately $-5$, $-45$, and $-70$ km\,s$^{-1}$, in Fig.\,\ref{fig_fuse01} indicated by dotted lines. 
Possibly, the two local Galactic components, in the 21\,cm-line emission data seen at $-11$ and $-17$ km\,s$^{-1}$, 
are smearing together into one single absorption component, given the fact the FUSE data has a spectral resolution of
only $\sim 20$ km\,s$^{-1}$. However, it is also possible that the narrow $-17$ km\,s$^{-1}$ 21\,cm-line 
component relates to gas that partly fills the $21\farcm0$ Green Bank beam, but is not present along the 
pencil-beam FUSE spectrum towards NGC\,3310. Since beam smearing effects can introduce enormous uncertainties 
when comparing 21\,cm-line data with FUV absorption line data for low-column density clouds, it is important to 
look for possibilities to study the structure of the neutral interstellar gas towards the Lockman Hole
based on the FUSE data alone. Unfortunately, the \ion{H}{i} Lyman series in the FUSE bandpass cannot be used for 
this purpose, since these lines are heavily saturated at this column density level.

The \ion{Ar}{i} $\lambda 1048.2$ (Fig.\,\ref{fig_fuse01}, third panel) indicates that the bulk of the neutral material
resides in the IVC component near $-45$ km\,s$^{-1}$. And also the H$_2$ absorption (lower-most
panel) suggests that most of the diffuse molecular gas is situated in the IV Arch rather than in the local 
Galactic component (see also \citet{ric03} for details on the H$_2$ spectrum). The best available tracer 
for the neutral gas in the FUSE bandpass is \ion{O}{i}. The \ion{O}{i} column density is tightly bound to the
\ion{H}{i} column density, since both neutral hydrogen and neutral oxygen have the same ionization potential 
($13.6$\,eV), and their abundance is coupled by charge-exchange reactions. In addition, oxygen is not strongly 
depleted in dust grains. Unfortunately, \ion{O}{i} absorption towards NGC\,3310 is saturated for all available lines 
at $\lambda\geq 930$\,\AA, and only the relatively weak \ion{O}{i} $\lambda 936.6$ can be used to roughly estimate 
the \ion{O}{i} column density. This line is shown in Fig.\,\ref{fig_fuse01}, second panel. As is clearly visible, 
the \ion{O}{i} absorption does not reach the zero-flux level in any of the three absorption components, 
which (together with the 
total width of the composite absorption structure) allows upper limits to be placed on the individual equivalent widths 
and column densities by modeling the component structure with Gaussian absorption components. Such a model
also has to account for the H$_2$ P(3) absorption that partly blends the local Galactic \ion{O}{i} absorption 
component near $+50$\,km\,s$^{-1}$ (in Fig.\,\ref{fig_fuse01} indicated with `b'). We have done such modeling for 
the \ion{O}{i} $\lambda 936.6$ line, and find that the \ion{O}{i} absorption in the local Galactic component 
at $-5$ km\,s$^{-1}$ must be $\leq 90$\,m\AA\, ($3\sigma$ upper limit), while for the two IVC
components at $-45$, and $-70$ km\,s$^{-1}$ we find equivalent width limits of $115$ and $85$\,m\AA.
Assuming that $b=10$\,km\,s$^{-1}$ (or greater), these equivalent width limits correspond to logarithmic column 
density limits of $\leq 16.02$, $\leq 16.57$, and $\leq 15.94$ for \ion{O}{i}, and $\leq 19.28$, $\leq 19.83$, 
and $\leq 19.20$ for \ion{H}{i}, respectively, if [O/H] is solar (log (O/H)$_{\sun}=-3.26$ \citep[][]{hol01}).
Thus, the FUSE data suggest that the \ion{H}{i} column density in the local Galactic gas towards NGC\,3310 is 
by a factor of $\sim 2$ lower than indicated by the beam-smeared 21\,cm-line data pointed in this direction,
while the 21\,cm-line matches the FUSE data for the two IVC components. However, unresolved 
sub-components (particular in the local Galactic component) may be present, in which case these limits could be incorrect.

What does the FUSE spectrum of NGC\,3310 tell us about the ionization conditions in the interstellar
gas in the direction of the Lockman Hole? As has been demonstrated in earlier studies \citep[][]{ric01b},
the \ion{P}{ii}/\ion{O}{i} column density ratio is a good indicator for the ionization structure in interstellar
clouds. \ion{P}{ii} has an ionization potential of 19.7 eV, and lives in regions where O and H are both neutral
and ionized. Also for phosphorus, dust depletion is not considered to be important. Thus, assuming an intrinsic
P/O ratio, the \ion{O}{i} and \ion{P}{ii} column densities can be used to determine lower limits for the 
ionization fraction, $f_{\rm ion}=$H$^+/$(H$^0+$H$^+$), in the weakly ionized gas phase (for ionization 
energies $<19.7$\,eV). The weak \ion{P}{ii} $\lambda 1152.8$ absorption (Fig.\,\ref{fig_fuse01}, fourth panel), can be reproduced 
by a three--component Gaussian fit, for which the three velocity components have been fixed at $-5$, $-45$, 
and $-70$ km\,s$^{-1}$. Using this method, we derive equivalent widths of $18\pm 4$ m\AA\, for the $-5$ km\,s$^{-1}$
component, $60\pm8$ m\AA\, for the $-55$ km\,s$^{-1}$ component, and $11\pm 4$ m\AA\, for the $-70$ km\,s$^{-1}$
component. Assuming a $b$ value of $10$ km\,s$^{-1}$, these equivalent widths correspond to column densities, 
$\log$ N$(\ion{P}{ii})$, of $12.85\pm0.10$, $13.51\pm0.09$, and $12.63\pm0.14$, respectively. Assuming that all 
three velocity components have solar phosphorus-to-oxygen abundance ratios \citep[see also][]{ric01a,ric01b},
the data for O and P yield $f_{\rm ion}\geq0.22$ and N(H)=N(\ion{H}{i})+N(H$^+)=8.66\times 10^{19}$cm$^{-2}$,
whereas no useful limits can be obtained for the local Galactic component and the second IVC component, since 
their P/O column density ratios would allow every degree of ionization (i.e., $f_{\rm ion}\geq0.0$). The FUSE data 
therefore indicates that approximately a fifth of the total hydrogen column density in the main IVC component
is in ionized form; very likely, the ionization in the local Galactic component is much higher, but the limitations 
for the FUSE \ion{O}{i} data unfortunately does not allow the derivation of a quantitative limit.

\subsubsection{Highly ionized gas}
\label{hig}
The FUSE data also yields information about the highly ionized gas phase for ionization energies $>100$ eV through 
the two resonance lines of five-times ionized oxygen, \ion{O}{vi}, situated at $1031.9$ and $1037.6$\,\AA. 
\ion{O}{vi} traces collisionally ionized, hot gas at temperatures $T\sim2\times10^5$\,K (thus lower than the X-ray 
emitting gas), or gas that is photoionized by a hard UV background (or a combination of both). In Fig.\,\ref{fig_fuse02} 
we have
plotted the \ion{O}{vi} $\lambda 1031.9$ absorption profile towards NGC\,3310 (the weaker \ion{O}{vi} $\lambda 1037.6$
cannot be used due to blending effects). The main analysis of the \ion{O}{vi} $\lambda 1031.9$ absorption towards 
NGC\,3310 has already been presented in \citet{wak03} and \citet{sav03}.
\ion{O}{vi} absorption is smeared over a broad range from $-200$ to $+90$ km\,s$^{-1}$, thus spanning a velocity 
range that covers all three velocity components seen in 21\,cm-line emission. The total integrated \ion{O}{vi} column 
density is $\log$ N$(\ion{O}{vi})=14.56\pm0.02$. This column density is slightly (but not significantly) higher 
than the \ion{O}{vi} column densities in other directions outside the Lockman Hole \citep[see][]{sav03}. 
This is not too surprising though, since the pathlength through the Lockman Hole is too short to contribute
significantly to the total \ion{O}{vi} column density at the expected low volume densities. Assuming that the 
fraction of oxygen residing in the \ion{O}{vi} phase is $\leq 0.2$, the total gas column density of this highly 
ionized gas phase is $\leq 3.3\times10^{18}$ cm$^{-2}$ for solar oxygen abundances \citep[][]{hol01}. The highly 
ionized gas component traced by \ion{O}{vi} therefore cannot contribute significantly to the total hydrogen gas column
density in direction of the Lockman Hole.

%________________________________________________________________
\section{Conclusion}
\label{conclusion}
The analysis performed shows that the main absorption of soft X-rays is due to the WNM. It turns out that a 
single-temperature plasma in the Milky Way halo is sufficient to reproduce the X-ray distribution
on the Lockman Hole field ($60^\circ \times 60^\circ$) very well. Hence, a patchy Milky Way halo is not likely.
Furthermore, the applied fit procedure (simultaneously for four ROSAT energy bands) is very sensitive to
temperature. We showed that the temperature of the LHB plasma is lower than the temperature of the halo plasma.
The X-ray/\ion{H}{i}-analysis of the Lockman Hole itself gives evidence that a large fraction of the attenuating 
ISM may reside in form of ionized hydrogen. 

While the FUSE data of NGC\,3310 impressively demonstrates the complex multiphase structure of the gas in the 
general direction of the Lockman Hole, a precise quantitative analysis of the gas properties turns out 
to be unfeasible due to the complexity of the velocity structure of the gas that cannot be resolved with the 
FUSE instrument. The data shows that the majority of the neutral and molecular gas is situated in the IVC component 
in the lower Milky Way halo rather than in the disk component. Due to the data limitations, we are not able to 
obtain precise ionization fractions, but the FUSE data of NGC\,3310 are at least {\em consistent} with the idea of 
ionized hydrogen in the Lockman Hole. While the ionization fraction is not well constrained by the FUSE data
the total amount of ionized gas is sufficiently high to overcome the apparent discrepancy between the \ion{H}{i}
and X-ray observations.

\begin{acknowledgements}
  MK and JK like to thank the Deutsches Zentrum f\"ur Luft- und
  Raumfahrt for financial support under grant No. 50 OR 0103.
\end{acknowledgements}

\bibliographystyle{aa}
\bibliography{ms3506}

\begin{thebibliography}{40}
\expandafter\ifx\csname natexlab\endcsname\relax\def\natexlab#1{#1}\fi

\bibitem[{{Barber} {et~al.}(1996){Barber}, {Roberts}, \& {Warwick}}]{bar96}
{Barber}, C.~R., {Roberts}, T.~P., \& {Warwick}, R.~S. 1996, \mnras, 282, 157

\bibitem[{{Dickey} \& {Lockman}(1990)}]{dic90}
{Dickey}, J.~M. \& {Lockman}, F.~J. 1990, \araa, 28, 215

\bibitem[{{Hartmann} \& {Burton}(1997)}]{har97}
{Hartmann}, D. \& {Burton}, W. 1997, Atlas of Galactic Neutral Hydrogen
  (Cambridge University Press)

\bibitem[{{Hartmann} {et~al.}(1996){Hartmann}, {Kalberla}, {Burton}, \&
  {Mebold}}]{har96}
{Hartmann}, D., {Kalberla}, P.~M.~W., {Burton}, W.~B., \& {Mebold}, U. 1996,
  \aaps, 119, 115

\bibitem[{{Hasinger} {et~al.}(2001){Hasinger}, {Altieri}, {Arnaud}, {Barcons},
  {Bergeron}, {Brunner}, {Dadina}, {Dennerl}, {Ferrando}, {Finoguenov},
  {Griffiths}, {Hashimoto}, {Jansen}, {Lumb}, {Mason}, {Mateos}, {McMahon},
  {Miyaji}, {Paerels}, {Page}, {Ptak}, {Sasseen}, {Schartel}, {Szokoly}, {Tr{\"
  u}mper}, {Turner}, {Warwick}, \& {Watson}}]{has01}
{Hasinger}, G., {Altieri}, B., {Arnaud}, M., {et~al.} 2001, \aap, 365, L45

\bibitem[{{Hausen} {et~al.}(2002){Hausen}, {Reynolds}, {Haffner}, \&
  {Tufte}}]{hau02}
{Hausen}, N.~R., {Reynolds}, R.~J., {Haffner}, L.~M., \& {Tufte}, S.~L. 2002,
  \apj, 565, 1060

\bibitem[{{Herbstmeier} {et~al.}(1993){Herbstmeier}, {Heithausen}, \&
  {Mebold}}]{her93a}
{Herbstmeier}, U., {Heithausen}, A., \& {Mebold}, U. 1993, \aap, 272, 514

\bibitem[{{Herbstmeier} {et~al.}(1995){Herbstmeier}, {Mebold}, {Snowden},
  {Hartmann}, {Butler Burton}, {Moritz}, {Kalberla}, \& {Egger}}]{her95}
{Herbstmeier}, U., {Mebold}, U., {Snowden}, S.~L., {et~al.} 1995, \aap, 298,
  606

\bibitem[{{Holweger}(2001)}]{hol01}
{Holweger}, H. 2001, Solar and Galactic Composition, AIP Conference, Proc. 598
  (ed. R.F. Wimmer-Schweingruber, (New York: American Institute of Physics),
  23)

\bibitem[{{Jahoda} {et~al.}(1990){Jahoda}, {Lockman}, \& {McCammon}}]{jah90}
{Jahoda}, K., {Lockman}, F.~J., \& {McCammon}, D. 1990, \apj, 354, 184

\bibitem[{{Kalberla} \& {Kerp}(1998)}]{kal98}
{Kalberla}, P.~M.~W. \& {Kerp}, J. 1998, \aap, 339, 745

\bibitem[{{Kalberla} {et~al.}(1980){Kalberla}, {Mebold}, \& {Velden}}]{kal80}
{Kalberla}, P.~M.~W., {Mebold}, U., \& {Velden}, L. 1980, \aaps, 39, 337

\bibitem[{{Kerp} {et~al.}(1999){Kerp}, {Burton}, {Egger}, {Freyberg},
  {Hartmann}, {Kalberla}, {Mebold}, \& {Pietz}}]{ker99}
{Kerp}, J., {Burton}, W.~B., {Egger}, R., {et~al.} 1999, \aap, 342, 213

\bibitem[{{Kerp} {et~al.}(1993){Kerp}, {Herbstmeier}, \& {Mebold}}]{ker93}
{Kerp}, J., {Herbstmeier}, U., \& {Mebold}, U. 1993, \aap, 268, L21

\bibitem[{{Kerp} {et~al.}(1996){Kerp}, {Mack}, {Egger}, {Pietz}, {Zimmer},
  {Mebold}, {Burton}, \& {Hartmann}}]{ker96}
{Kerp}, J., {Mack}, K.-H., {Egger}, R., {et~al.} 1996, \aap, 312, 67

\bibitem[{{Kuntz} \& {Snowden}(2000)}]{kun00}
{Kuntz}, K.~D. \& {Snowden}, S.~L. 2000, \apj, 543, 195

\bibitem[{{Lockman} {et~al.}(1986){Lockman}, {Jahoda}, \& {McCammon}}]{loc86}
{Lockman}, F.~J., {Jahoda}, K., \& {McCammon}, D. 1986, \apj, 302, 432

\bibitem[{{McCammon} \& {Sanders}(1990)}]{mcc90}
{McCammon}, D. \& {Sanders}, W.~T. 1990, \araa, 28, 657

\bibitem[{{Morrison} \& {McCammon}(1983)}]{mor83}
{Morrison}, R. \& {McCammon}, D. 1983, \apj, 270, 119

\bibitem[{{Murphy} {et~al.}(1996){Murphy}, {Lockman}, {Laor}, \&
  {Elvis}}]{mur96}
{Murphy}, E.~M., {Lockman}, F.~J., {Laor}, A., \& {Elvis}, M. 1996, \apjs, 105,
  369

\bibitem[{{Pietz} {et~al.}(1998){Pietz}, {Kerp}, {Kalberla}, {Burton},
  {Hartmann}, \& {Mebold}}]{pie98}
{Pietz}, J., {Kerp}, J., {Kalberla}, P.~M.~W., {et~al.} 1998, \aap, 332, 55

\bibitem[{{Plucinsky} {et~al.}(1993){Plucinsky}, {Snowden}, {Briel},
  {Hasinger}, \& {Pfeffermann}}]{plu93}
{Plucinsky}, P.~P., {Snowden}, S.~L., {Briel}, U.~G., {Hasinger}, G., \&
  {Pfeffermann}, E. 1993, \apj, 418, 519

\bibitem[{{Richter} {et~al.}(2001{\natexlab{a}}){Richter}, {Savage}, {Wakker},
  {Sembach}, \& {Kalberla}}]{ric01a}
{Richter}, P., {Savage}, B.~D., {Wakker}, B.~P., {Sembach}, K.~R., \&
  {Kalberla}, P.~M.~W. 2001{\natexlab{a}}, \apj, 549, 281

\bibitem[{{Richter} {et~al.}(2001{\natexlab{b}}){Richter}, {Sembach}, {Wakker},
  {Savage}, {Tripp}, {Murphy}, {Kalberla}, \& {Jenkins}}]{ric01b}
{Richter}, P., {Sembach}, K.~R., {Wakker}, B.~P., {et~al.} 2001{\natexlab{b}},
  \apj, 559, 318

\bibitem[{{Richter} {et~al.}(2003){Richter}, {Wakker}, {Savage}, \&
  {Sembach}}]{ric03}
{Richter}, P., {Wakker}, B.~P., {Savage}, B.~D., \& {Sembach}, K.~R. 2003,
  \apj, 586, 230

\bibitem[{{Savage} {et~al.}(1997){Savage}, {Sembach}, \& {Lu}}]{sav97}
{Savage}, B.~D., {Sembach}, K.~R., \& {Lu}, L. 1997, \aj, 113, 2158

\bibitem[{{Savage} {et~al.}(2003){Savage}, {Sembach}, {Wakker}, {Richter},
  {Meade}, {Jenkins}, {Shull}, {Moos}, \& {Sonneborn}}]{sav03}
{Savage}, B.~D., {Sembach}, K.~R., {Wakker}, B.~P., {et~al.} 2003, \apjs, in
  press, astroph/0208140

\bibitem[{{Sfeir} {et~al.}(1999){Sfeir}, {Lallement}, {Crifo}, \&
  {Welsh}}]{sfe99}
{Sfeir}, D.~M., {Lallement}, R., {Crifo}, F., \& {Welsh}, B.~Y. 1999, \aap,
  346, 785

\bibitem[{{Snowden} {et~al.}(1998){Snowden}, {Egger}, {Finkbeiner}, {Freyberg},
  \& {Plucinsky}}]{sno98}
{Snowden}, S.~L., {Egger}, R., {Finkbeiner}, D.~P., {Freyberg}, M.~J., \&
  {Plucinsky}, P.~P. 1998, \apj, 493, 715

\bibitem[{{Snowden} {et~al.}(1997){Snowden}, {Egger}, {Freyberg}, {McCammon},
  {Plucinsky}, {Sanders}, {Schmitt}, {Truemper}, \& {Voges}}]{sno97}
{Snowden}, S.~L., {Egger}, R., {Freyberg}, M.~J., {et~al.} 1997, \apj, 485, 125

\bibitem[{{Snowden} \& {Freyberg}(1993)}]{sno93}
{Snowden}, S.~L. \& {Freyberg}, M.~J. 1993, \apj, 404, 403

\bibitem[{{Snowden} {et~al.}(2000){Snowden}, {Freyberg}, {Kuntz}, \&
  {Sanders}}]{sno00}
{Snowden}, S.~L., {Freyberg}, M.~J., {Kuntz}, K.~D., \& {Sanders}, W.~T. 2000,
  \apjs, 128, 171

\bibitem[{{Snowden} {et~al.}(1995){Snowden}, {Freyberg}, {Plucinsky},
  {Schmitt}, {Truemper}, {Voges}, {Edgar}, {McCammon}, \& {Sanders}}]{sno95}
{Snowden}, S.~L., {Freyberg}, M.~J., {Plucinsky}, P.~P., {et~al.} 1995, \apj,
  454, 643

\bibitem[{{Snowden} {et~al.}(1994{\natexlab{a}}){Snowden}, {Hasinger},
  {Jahoda}, {Lockman}, {McCammon}, \& {Sanders}}]{sno94a}
{Snowden}, S.~L., {Hasinger}, G., {Jahoda}, K., {et~al.} 1994{\natexlab{a}},
  \apj, 430, 601

\bibitem[{{Snowden} {et~al.}(1994{\natexlab{b}}){Snowden}, {McCammon},
  {Burrows}, \& {Mendenhall}}]{sno94}
{Snowden}, S.~L., {McCammon}, D., {Burrows}, D.~N., \& {Mendenhall}, J.~A.
  1994{\natexlab{b}}, \apj, 424, 714

\bibitem[{{Snowden} {et~al.}(1991){Snowden}, {Mebold}, {Hirth}, {Herbstmeier},
  \& {Schmitt}}]{sno91}
{Snowden}, S.~L., {Mebold}, U., {Hirth}, W., {Herbstmeier}, U., \& {Schmitt},
  J.~H.~M. 1991, Science, 252, 1529

\bibitem[{{Tozzi} {et~al.}(2001){Tozzi}, {Rosati}, {Nonino}, {Bergeron},
  {Borgani}, {Gilli}, {Gilmozzi}, {Hasinger}, {Grogin}, {Kewley}, {Koekemoer},
  {Norman}, {Schreier}, {Szokoly}, {Wang}, {Zheng}, {Zirm}, \&
  {Giacconi}}]{toz01}
{Tozzi}, P., {Rosati}, P., {Nonino}, M., {et~al.} 2001, \apj, 562, 42

\bibitem[{{Wakker}(2001)}]{wak01}
{Wakker}, B.~P. 2001, \apjs, 136, 463

\bibitem[{{Wakker} {et~al.}(2003){Wakker}, {Savage}, {Sembach}, {Richter},
  {Meade}, {Jenkins}, \& {Shull}}]{wak03}
{Wakker}, B.~P., {Savage}, B.~D., {Sembach}, K.~R., {et~al.} 2003, \apjs, in
  press, astroph/0208009

\bibitem[{{Wei{\ss}} {et~al.}(1999){Wei{\ss}}, {Heithausen}, {Herbstmeier}, \&
  {Mebold}}]{wei99}
{Wei{\ss}}, A., {Heithausen}, A., {Herbstmeier}, U., \& {Mebold}, U. 1999,
  \aap, 344, 955

\end{thebibliography}

\end{document}